\documentclass[conference]{IEEEtran}
\IEEEoverridecommandlockouts

\PassOptionsToPackage{draft}{hyperref}

\ifCLASSINFOpdf
\else
\fi

\usepackage{booktabs}
\usepackage[utf8]{inputenc}
\usepackage{amsmath}
\usepackage{amsfonts}
\usepackage{amssymb}
\usepackage{graphicx}
\usepackage{listings}
\usepackage{caption,subcaption}
\usepackage{algorithm}
\usepackage[noend]{algpseudocode}
\usepackage[utf8]{inputenc}
\usepackage[english]{babel}
\usepackage{tabularx}
\usepackage{paralist}
\usepackage{multirow}
\usepackage[T1]{fontenc}
\usepackage[scaled=.8]{beramono}
\def\BibTeX{{\rm B\kern-.05em{\sc i\kern-.025em b}\kern-.08em
    T\kern-.1667em\lower.7ex\hbox{E}\kern-.125emX}}

\newcommand{\tool}{\textsc{JSNeat}\xspace}
\newcolumntype{L}[1]{>{\raggedright\arraybackslash}p{#1}}

\usepackage{amsthm}

\usepackage{xcolor}
\usepackage{color, colortbl}
\usepackage{hyperref}

\let\labelindent\relax
\usepackage{enumitem}

\usepackage{balance}

\newtheorem{definition}{Definition}

\usepackage{xspace}

\newcommand{\eg}{\hbox{\emph{e.g.,}}\xspace}
\newcommand{\ie}{\hbox{\emph{i.e.,}}\xspace}

\definecolor{lightgray}{rgb}{.9,.9,.9}
\definecolor{darkgray}{rgb}{.4,.4,.4}
\definecolor{purple}{rgb}{0.65, 0.12, 0.82}

\lstdefinelanguage{JavaScript}{
  keywords={typeof, new, true, false, catch, function, return, null, catch, switch, var, if, in, while, do, else, case, break},
  keywordstyle=\color{blue}\bfseries,
  ndkeywords={class, export, boolean, throw, implements, import, this},
  ndkeywordstyle=\color{darkgray}\bfseries,
  identifierstyle=\color{black},
  sensitive=false,
  comment=[l]{//},
  morecomment=[s]{/*}{*/},
  commentstyle=\color{purple}\ttfamily,
  stringstyle=\color{red}\ttfamily,
  morestring=[b]',
  morestring=[b]"
}

\begin{document}
%

\title{Recovering Variable Names for Minified Code\\ with Usage Contexts}





%


\author{\IEEEauthorblockN{Hieu Tran\IEEEauthorrefmark{1},
Ngoc Tran\IEEEauthorrefmark{1},
Son Nguyen\IEEEauthorrefmark{1},
Hoan Nguyen\IEEEauthorrefmark{2}, and
Tien N. Nguyen\IEEEauthorrefmark{1}}
  \IEEEauthorblockA{\IEEEauthorrefmark{1}Computer Science Department, The University of Texas at Dallas, USA,\\ Email: \{trunghieu.tran,nmt140230,sonnguyen,tien.n.nguyen\}@utdallas.edu}
  \IEEEauthorblockA{\IEEEauthorrefmark{2}Computer Science Department,
Iowa State University, USA, Email: hoan@iastate.edu}}


\maketitle

\begin{abstract}

To avoid the exposure of original source code in a Web application,
the variable names in JS code deployed in the wild are often replaced
by short, meaningless names, thus making the code extremely difficult
to manually understand and analysis. This paper presents {\tool}, an
information retrieval (IR)-based approach to recover the variable
names in minified JS code. {\tool} follows a data-driven approach to
recover names by searching for them in a large corpus of open-source
JS code. We use three types of contexts to match a variable in given
minified code against the corpus including the context of the properties
and roles of the variable, the context of that variable and relations
with other variables under recovery, and the context of the task of
the function to which the variable contributes. We performed several
empirical experiments to evaluate {\tool} on the dataset of more than
322K JS files with 1M functions, and 3.5M variables with 176K unique
variable names. We found that {\tool} achieves a high accuracy of
69.1\%, which is the relative improvements of 66.1\% and 43\% over two
state-of-the-art approaches JSNice and JSNaughty, respectively. The
time to recover for a file or a variable with {\tool} is twice as
fast as with JSNice and 4x as fast as with JNaughty, respectively.

\end{abstract}


\begin{IEEEkeywords}
Minified JS Code, Variable Name Recovery, Naturalness of Code, Usage Contexts.
\end{IEEEkeywords}

%

\IEEEpeerreviewmaketitle

\section{Introduction}
\label{intro-section}


Software developers have to spend a
significant portion~of their efforts in comprehending the code. An
important aspect of program understanding
is the names of the identifiers~used in the source
code~\cite{sutton-fse15}. Meaningful identifiers help developers
tremendously in quickly grasping the essence of the~code. Thus,
naming conventions are strongly emphasized on prescribing how to
choose meaningful variable names in coding
standards~\cite{barr-codeconvention-fse14}.
These principles also apply to Web development.

Web technologies and programming languages require the exposure of
source code to Web browsers in the client side to be executed
there. To avoid such exposure, the source code such as JavaScript (JS)
files are often obfuscated in which the variable names are minified,
\ie the variable names are replaced with short, opaque, and
meaningless names. The intention has two folds. First, it makes the JS
files smaller and thus is quickly loaded for better performance.
Second, minification diminishes code readability to hide business
logics from the readers, while maintaining the program semantics.
%
%

Due to those reasons, there is a natural need to automatically recover
the minified code with meaningful variable names. When the original
code is not available, with such recovery, the minified JS code will
be made accessible for code~compre-\\hension as well as other maintenance
activities such as~code review, reuse, analysis, and
enhancement. Recognizing that need, researchers have been introducing
the automatically recovering tools for variable names in JS code.
JSNice~\cite{JSNice2015} is an automatic variable name recovery
approach that represents the program properties and relations among
program entities in a JS code as dependence graphs.
It leverages advanced machine learning (ML) to recover missing
variable names.
%
Using also ML, JSNaughty~\cite{JSNaughty2017}
formulates the variable name recovery problem for JS code as a
statistical machine translation from minified code to the
recovered code. Despite of their successes, both approaches still
suffer low accuracy and scalability issues with the use of
computationally expensive ML algorithms.

In this work, we present {\tool}, a data-driven, information retrieval
(IR)-based approach to automatically recover variable~name  for minified
JS code. The recovered names for variables must be natural in the
context of the code and follow naming conventions. Thus, we conform
{\tool} in a data-driven direction, in which we aim to search for the
name recovered for a minified variable in a large corpus of
open-source JS code. We conjecture that meaningful, natural names of
minified variables could be seen before in such corpus.
Our key idea is to {\em utilize the contexts for the variables in
  source code to search for its name}.
For the minified variables in a given a minified JS code, {\tool} aims
to match their contexts against the contexts in the corpus. If two
contexts of two variables are similar, they should be named similarly,
thus the variable's name in the corpus with the matched contexts
should be a candidate name for the corresponding minified~variable.

For a given minified variable $v$, we model three types of contexts.
First, {\em the name of a variable should be affected by its own
  properties and roles in the source code}. For properties, if $v$
accesses to a method $m$ or a field $f$ in the code, then {\em the
  recovered name for $v$ should be naturally compatible with the names
  of the method and the field}. For example, the variable named
\texttt{dataTransfer} is the receiver of the method call
\texttt{getData()}.
%
However, a variable that calls the method \texttt{getData()} cannot
be randomly named in a regular program.
For a role, if $v$ is used an argument of a method call $m$, the data
type of $v$ must be compatible with $m$, thus,
their names should also naturally be in conformance with one
another. For example, in \texttt{JQuery.trigger(...)}, the first
argument is either an event or an event type, thus, the name of the
first argument should be consistent with a direct object of the verb
\texttt{trigger}.

Second, {\em the name of a variable should be affected by the names of
  variables co-occurring in the same function}. Several variables are
used together to contribute to the current task of the function, thus,
their names are naturally in concordance with one another. Observing
some variable names, {\tool} could predict the names for other
co-occurring variables. It takes into account the naming of multiple
variables at once considering such co-appearances. Finally, {\em the
  third type of context is the current task of the function to which
  the variable belongs and contributes}. The names of the variable
should be relevant to the common task/purpose of the function. For
example, the variables in a function \texttt{getClipboardContent}
should serve the task of getting the content from the clipboard and
have the names closely relevant to that task.

We combine the above contexts in the {\tool} tool.  We built a
database to store the information about the variable names and the
contexts extracted from a large corpus of open-source JS code. To
recover the names for a given minified JS file, we use {\tool} with
three types of contexts to search for and rank the candidate names. We
performed several experiments to evaluate {\tool} on the dataset of
322K JS files with almost 1M functions, and 3.5M variables with 176K
unique variable names. {\tool} achieves high accuracy of {\em 69.1\%},
which is the relative improvements of {\em 66.1\%} and {\em 43\%} over
two state-of-the-art approaches JSNice~\cite{JSNice2015} and
JSNaughty~\cite{JSNaughty2017}, respectively. A high percentage ({\em
  29.4\%}) of variables is recovered only by {\tool}, while 4.7\% and
3.6\% of variables are recovered only by JSNaughty and JSNice,
respectively. We reported that the time to recover for a file or for a
variable with {\tool} is {\em twice as fast as with JSNice and 4x as
  fast as with JNaughty}, respectively. Importantly, {\em {\tool}'s
  training time is 4x faster than JSNice and 6x faster than
  JSNaughty}.
%
This paper makes the following contributions:

{\bf 1.} {\tool}: an IR-based, data-driven approach to~recover
variable names for minified JS code using 3 types of contexts;

{\bf 2.} An extensive comparative evaluation and analysis
 on {\tool}'s accuracy and running time to show that it outperforms
 the state-of-the-art approaches (See results in
 a~website~\cite{jsneat-website});

{\bf 3.} A novel formulation of variable name recovery problem in
minified JS code as an Information Retrieval problem.












\section{Motivation and Approach Overview}
\label{example_section}

Figures~\ref{example_org} and~\ref{example_sim} show the
original and minified versions~of the JS function
\texttt{get\-Clip\-board\-Content} in the \texttt{vue-medium-ed\-itor}
project.
%
The function is to retrieve the content of~the~clipboard. In the minified code, all local variables~are
randomly renamed with short and meaningless names, \eg
\texttt{data\-Transfer} becomes \texttt{r}, \texttt{data} becomes
\texttt{n}, by a minification tool, \eg UglifyJS \cite{uglify}. This
makes developers difficult to comprehend~it.



\begin{figure}[t]
\begin{center}
\begin{lstlisting}[captionpos=b, numbers=left, stepnumber=1, numbersep=-6pt,
    frame=single, xleftmargin=4pt, xrightmargin=4pt]
  function getClipboardContent(event, win, doc) {
    var dataTransfer = event.clipboardData || win.clipboardData || doc.dataTransfer,
        data = {};
    if (!dataTransfer) return data;
    if (dataTransfer.getData) {
       var legacyText=dataTransfer.getData('Text');
       if (legacyText && legacyText.length > 0) {
          data['text/plain'] = legacyText;
       }
    }
    if (dataTransfer.types) {
       for (var i = 0; i < dataTransfer.types.length; i++) {
         var contentType = dataTransfer.types[i];
         data[contentType] = dataTransfer.getData(contentType);
       }
    }
    return data;
  }
\end{lstlisting}
\caption{An Original Code from a Project in GitHub}
\label{example_org}
\end{center}
\end{figure}

\begin{figure}[t]
\begin{center}
\begin{lstlisting}[captionpos=b, numbers=left, stepnumber=1, numbersep=-6pt,
    frame=single, xleftmargin=4pt, xrightmargin=4pt]
  function getClipboardContent(t, a, e) {
    var r = t.clipboardData ||a.clipboardData ||e.dataTransfer,
          n = {};
    if (!r) return n;
    if (r.getData) {
       var i = r.getData("Text");
       if (i && i.length > 0) {
          n["text/plain"] = i;
       }
    }
    if (r.types) {
       for (var p = 0; p < r.types.length; p++) {
         var f = r.types[f];
         n[f] = r.getData(f);
       }
    }
    return n;
  }
\end{lstlisting}
\caption{The Minified Code for the Code in Figure~\ref{example_org}}
\label{example_sim}
\end{center}
\end{figure}

Our goal is to assign meaningful names for the variables in the
minified code.
The name chosen for a variable in the code should be {\em natural}
(unsurprising) in the context~\cite{JSNaughty2017} and follow naming
conventions~\cite{barr-codeconvention-fse14}, so that the de-minified
code becomes easy to understand for developers.

To achieve this goal, we conjecture that the meaningful names of
minified variables could be observed in a large corpus of existing
source code. This motivates us to conform our approach to a {\em
  data-driven direction}, where we learn the names from original source
code to recover the names for variables in the minified code.
Indeed, for the minified code in Figure~\ref{example_sim}, all
original names are found in our experimental dataset that contains
322K JS files collected from 12K GitHub~projects.

\subsection{Observations}

The name recovering process of variables in
minified code is affected by multiple factors. Let us illustrate these
factors through the following observations:

\textbf{O1}. \textit{Each individual variable has certain properties
  and plays particular roles in the code. Thus, the name of a variable
  is intuitively affected by its properties and roles}. The properties
are the method calls or field accesses to which a variable of a certain
type can access.
%
If a method is called or a field is accessed
by a variable, the name of the variable should be compatible with the
method's or the field's name.
%
%
%
For example, in our experimental dataset,
the number of candidates that can call method \texttt{getData()}
(lines 6 and 14) is only 7 out of 31 variables names found in a
function named \texttt{getClipboardContent}.
%
%
Such number is down to a \textit{single} candidate if we additionally
consider that it can also access the fields \texttt{getData} (line 5)
and \texttt{types} (line 11). Thus, \texttt{r} could be named
as \texttt{dataTransfer}, which is the same name in the
original code in Figure~\ref{example_org}. For the variable \texttt{f}
that is created and assigned as an element of the array
\texttt{types[]} at line 13, there are 4 candidates for such
variable that can be used as an argument of the method named
\texttt{getData()} (line~14). The number of candidates for \texttt{i},
which is the returned result of the call to \texttt{getData()} (line
6) and also has a field with the name \texttt{length} (line 7), is
only 7.

\textbf{O2}. \textit{In a function, a variable might collaborate with
  other variables to implement the function. Consequently, the
  recovering name for a variable might be influenced by the name of
  others}. Intuitively, since the variables are used together, their
names are often consistent with each other to achieve the
common task in the function.
In the example, in 28 possible pairs of candidates for \texttt{i} (7
candidates) and \texttt{f} (4 candidates), there are only 2 pairs of
candidates that are used to name two variables in the same function
in our dataset. One of them is the correct pair, which is
\texttt{legacyText} and \texttt{contentType}.
%

\textbf{O3.} \textit{Within a function, e.g.,
  \texttt{getClipboardContent}, a variable name, e.g.,
  \texttt{contentType} is affected by the specific task of the
  function that is described by the function's
  name~\cite{sutton-fse15}}.~This is intuitive because the names of
variables are often~relevant~to the task that the variables are used
in the code~to~achieve. Such task is typically described with a
succinct function~name.
In Figure~\ref{example_org}, the task of~the function is to get the
clipboard's content, thus, it is named
\texttt{get\-ClipboardContent}. In our~dataset, there are 31
names being used to specify~the variables in function
\texttt{getClipboardContent}, \eg \texttt{data},
\texttt{dataTransfer}, \texttt{contentType}.
Meanwhile, the variable names \texttt{students}
or \texttt{salary} have never been used in the function with that
name.


Overall, these observations indicate that the names of the variables
in a particular function not only depend on {\em the task} in which
the variable is used to implement (called {\bf task-specific context}), but
their names are also affected by {\em their own properties and roles} in the
code (called {\bf single-variable usage context}) and on the {\em names of the
other variables} in the same function (called {\bf multiple-variable usage
  context}).

\subsection{Approach Overview}

From the observations, we propose an IR-based, data-driven approach to
recover the variables' names in a minified JS code based
on the contextual information
including \textit{single-variable usage context} (SVC),
\textit{multiple-variable usage context} (MVC), and
\textit{task-specific context} (TSC).
We initially construct a database to store
the variables' names and the corresponding context information
extracted from a large corpus of JS code.
To recover the names, given a minified JS code, we first use
the \textit{SVC} and \textit{TSC} information to find in our
database the candidate names for each variable. Then, these candidates
for each variable are ranked by the likelihood that they are used
along with the candidates of other variables, in order to name the
variables in the same function by using~\textit{MVC}.

\section{Single-Variable Usage Context (SVC)}






This section presents the single-variable context that we
use in the name recovery process. The intuition for this context is
that to recover the name of a variable, one could use its own context
based on its own {\bf properties} and {\bf roles} in the code.

By properties of a variable, we refer to the methods or fields to
which a variable of certain type can access.
In the minified code, the names of the called methods and accessed fields are 
not minified. Thus, it could play the role of the pivots~in recovering the 
variables' names.
Importantly, due to nature~of naming, {\em the name of~the variable
should be compatible with the name of the method being called or the
name of the field being accessed}. Thus, they provide hints on
the names of~the variables.
For example, in Figure~\ref{example_sim}, the only candidate
that calls the method \texttt{getData} and accesses the
field \texttt{types} is \texttt{dataTransfer}.
%
Those names are compatible with each other. To learn compatible names,
we follow a data-driven approach by learning from a large corpus of
non-minified JS code.


By the role of a variable, we refer to its usage context with the
method calls or field accesses that were not minified. For example, a
variable could be an argument of a method call, or a variable could
be assigned with the returned value from a method~call or field access.
%
The name of a variable used as an argument is often compatible with
its type/role and, thus, coupled with the name of the method call itself.
%
On line 14 of Figure~\ref{example_org}, the argument
\texttt{contentType} is in conformance with the method name
\texttt{getData}. Thus, it helps recover the name of the
minified variable \texttt{f} on line 14 of Figure~\ref{example_sim}.
Similarly, the name of a variable receiving the returned value of
a method call or a field access should conform with the name of the
method or field. Such conformance can be learned from
a large corpus of non-minified code, and helps recover variable~names.



\subsection{Property and Role Relations}

To realize the single-variable context for name recovery with
properties and roles of a variable, we define two key relations: 
{\em Property} and {\em Role}. Those relations form the
single-variable usage context for name recovery.



\begin{definition}{\bf [Property Relation (PropRel)]}
Property relation represents the relationships between a variable and
its fields or methods to which the variable can access or call.
\end{definition}


A property relation between a variable \texttt{v} and its
property \texttt{p} is denoted by a triple $(v, p, t)$, 
where \texttt{t} is the type of relation, which can be 
either \texttt{fieldAccess} or \texttt{methodCall}.
In Figure~\ref{example_sim}, a set of property relations for \texttt{r}
includes $( \texttt{r}, \texttt{types}, \texttt{fieldAccess})$,
$( \texttt{r}, \texttt{getData}, \texttt{fieldAccess})$,
$( \texttt{r}, \texttt{getData()}, \texttt{methodCall})$.  


%

\begin{definition}{\bf [Role Relation (RoleRel)]}
  Role relation represents the relationships between a variable and
  the method calls or field accesses in its usages.
\end{definition}




Since the names of methods or fields are not minified, we consider
them as the pivots in the usage context for recovering names of the
minified variables.
We focus on the roles of a variable used {\em as an argument in a method call}
 or {\em receiving the value returned by a method call or field access}.
If we have $o.m(...,v,...)$, $v = o.m(...)$, or $v = o.f$,
then there exist
the role relations between $v$ and $m$, and between $v$ and $f$. The
rationale is that the names of $m$ and its argument are often in
conformance with each other, \eg
\texttt{getData(contentType)}. Similar rationale is applied to the
above assignments to $v$.


A role relation between a variable \texttt{v} and a field/method
\texttt{p} is denoted by a triple $(v, p, t)$, where
\texttt{t} is the type of role relation. A role relation could
be either \texttt{argument} or \texttt{assignment}.




\subsection{Graph Representation of Single-Variable Context}
\label{graph_section}


%

\begin{definition}{\bf [Relation Graph]}
A relation graph (RG) for a variable $v$ is a directed graph
in the shape of a star to represent the single-variable usage context
of $v$ with regard to its property and role relations with 
the fields and methods in its usage.
The center vertex of the RG represents the variable. The other vertices
represent the methods/fields in method calls or field accesses, respectively, and are labeled with their names. 
Edges represent relations and are labeled with relation types. 
\end{definition}



\begin{figure}[t]
	\begin{center}
		\includegraphics[width=.7\columnwidth]{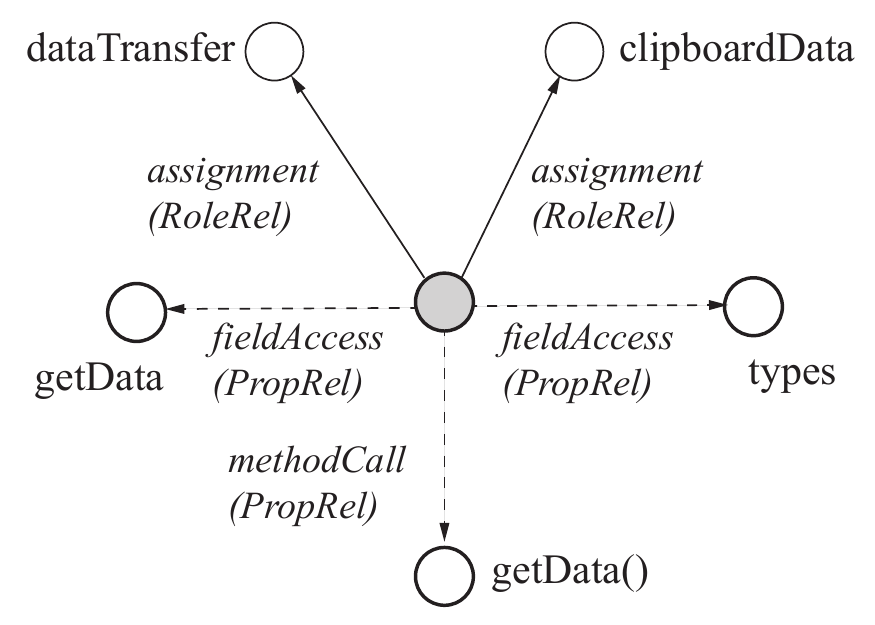}
		\caption{The Relation Graph of Variable $r$ in Figure~\ref{example_sim}}
		\label{SG_sample_ano}
	\end{center}
\end{figure}

Figure~\ref{SG_sample_ano} shows the relation graph of the variable
\texttt{r}, which includes a set of property relations:
$( \texttt{r}, \texttt{types}, \texttt{fieldAccess})$, $( \texttt{r},
\texttt{getData}, \texttt{fieldAccess})$, $( \texttt{r},
\texttt{getData()}, \texttt{methodCall})$,
and a set of role relations: $( \texttt{r}, \texttt{clipboardData},
\texttt{assignment})$, $( \texttt{r}, \texttt{dataTransfer},
\texttt{assignment})$ in our example.




%

\subsection{Deriving Candidate Names Using Single-Variable Context}
\label{derive_single_var_section}

Let us explain how we use the relation graphs to derive
the ranked list of candidate names for a minified~variable.



Our idea is that if two variables have the same/similar contexts, they
are often named similarly.
Given a minified function $f$, we first parse $f$ to produce a
relation graph $G_v$ for each variable $v$.
We then search for the single-variable contexts that are matched
with the context of $v$ within a dataset $\mathcal{G}$ of the relation
graphs built from a large corpus of open-source projects (We will
explain how to build the dataset later).


\begin{definition}{\bf [Single-Variable Context Matching]}
  Two single-variable usage contexts for a minified variable $v$ and a
  variable $v'$ in the dataset are considered to be matched if and
  only if their corresponding relation graphs are matched.
\end{definition}

\begin{definition}{\bf [Relation Graph Matching]}
A relation graph $G_v$ of a minified variable $v$ is considered as
matched with relation graph $G_{v'}$ of $v'$ in $\mathcal{G}$ if and
only if their graph matching score is equal or greater than
a threshold $\varphi$.
\end{definition}

Since RGs all have star shape, matching graphs can be done by matching
their sets of edges.

\begin{definition}{\bf [Relation Graph Matching Score]}
The graph matching score $\rho(G_v, G_{v'})$ between $G_v$ of a
minified variable $v$ and a relation graph $G_{v'}$ of $v'$ is
computed as the percentage of the number of edges in $G_v$ found in
$G_{v'}$.
\end{definition}

A variable name might appear in multiple functions, so it might have
multiple relation graphs. Thus, we define a name matching score
considering all of those functions as follows.

\begin{definition}{\bf [Single-Variable Score]}
Single-variable score represents how well name $vn$ can be
used for a minified name $v$ and is computed based on graph matching score:

\begin{equation}
\label{eq:sgcs}
SC_{v, vn} = \max_{G_{v'} \in \mathcal{G}_{vn}} \rho({G_v, G_{v'})}
\end{equation}

where
$\mathcal{G}_{vn}$ is the set of relation graphs of name $vn$ that match $G_v$.
$\mathcal{G}_{vn}$ represents multiple usages of the name $vn$.
\end{definition}





If a match is found, the name $vn$ of the variable $v'$ in the matched
relation graph in $\mathcal{G}$ is considered as a candidate name for
$v$.
There might exist many candidate names having similar contexts with
the context of a variable $v$. The candidate names are ranked
based on their name matching scores.
%
%
The higher the name matching score, the higher the confidence of our
model in using the name $vn$ of $v'$ for $v$.

%




For example, when recovering the variable \texttt{r} in
Figure~\ref{example_sim}, by searching on the dataset, we found that
the variables named \texttt{dataTransfer} and \texttt{dataObj} have
the relation graphs matching with that of \texttt{r}. This implies
that these variable names have been used in the past and they have
similar SVC contexts~with \texttt{r}. Therefore, \texttt{r}
could be recovered as \texttt{dataTransfer} or \texttt{dataObj}.

\section{Multiple-variable Usage Context (MVC)}

Let us present how we define and use multiple-variable usage
context. To achieve a specific task,
developers use one or multiple variables in their code.
Because the variables all play their roles in the code, their names
are often relevant and consistent with one another in order to achieve
the common task in the function to which they belong. For example, in
Figure~\ref{example_org}, the
variables \texttt{dataTransfer}, \texttt{contentType},
and \texttt{data} serve their roles in the task to retrieve the
content of a clipboard, and their names are naturally consistent with
one another with regard to that task.
In name recovery, we utilize such co-occurrences of variable names to
recover the name for one variable while another one was recovered with
the co-occurring name if the contexts of two variables allow.

\subsection{Multiple-Variable Score}
To formulate the co-occurrence of variable names, we define the {\em
  association score} for a set of variable names, which represents how
likely those names appear together in a function.
Assume that we have a set of $n$ variable names, and the name of
$i^{th}$ variable is $vn_i$. The association score for a set $S$ of
\textit{n} names $(vn_1, vn_2, ..., vn_n)$ is computed as follows.
\begin{equation}
\label{eq:assocN}
assoc(S)=\frac{N_{vn_1 \cap vn_2 \cap ... \cap vn_n}}{N_{vn_1 \cup vn_2 \cup ... \cup vn_n}}
\end{equation}




\noindent where
$N_{vn_1 \cap vnn_2 \cap ... \cap vn_n}$ is the number of functions that contain all the names $(vn_1, vn_2, ..., vn_n)$ in the corpus. 
$N_{vn_1 \cup vn_2 \cup ... \cup vn_n}$ is the number of functions that contain at least one of the names $(vn_1, vn_2, ..., vn_n)$ in the corpus.







For a set of $n$ recovered names, we define a {\bf Multiple-variable score}
($MC$) that represents the likelihood of those variable names to be
assigned to the variables based on~MVC. $MC$ can be computed using the
association score. However, due to the fact that not all possible sets
of $n$ names appear together in a corpus, we compute $MC$ based on the
associations of all subsets of the size $J$ with $J$ $\leq$~$n$, as~follows:
\begin{equation}
\label{confidentScore}
MC_{vn_1, vn_2, ..., vn_n} = \frac{\sum_{i=1}^{n_{subJ}} assoc(S_i)}{n_{subJ}}
\end{equation}
where $S_i$ is a subset of size $J$ of $(vn_1, vn_2, ..., vn_n)$;
$n_{subJ}$ is the number of such subset; and $assoc(S_i)$ is the
association score of all variable names in set $S_i$ (computed by
Formula~\ref{eq:assocN}).

\subsection{Deriving Candidate Names Using Multi-Variable Context}
\label{multivar-algo-section}

This section presents our algorithm to derive  
candidate names using multiple-variable usage context.
The algorithm takes as input a set of minified variables in a JS code
in which each variable has a set of candidate names (derived using the
single-variable context as explained in
Section~\ref{derive_single_var_section} or using task-specific context
in Section~\ref{task_section}).
%
%
The output is the ranked list of the results with associated
scores. Each result is a set of the recovered names for all of the given
variables.


\subsubsection{Design Strategies}

%

%

%

In developing our algorithm, we face three key challenges. First, each
variable might have a large number of candidates, thus, there are an
exponential combination among variables' names. {\em How would we
deal~with such complexity to make our algorithm scale?}  Second, {\em
given a set of minified variables needed to be recovered, which one
should {\tool} start}? This is important since if the algorithm
does not recover well the first variable, this would affect much to
the accuracy of recovering the next variables.
%
Finally, {\em in which recovery order for the variables in a function
would it be beneficial from multiple-variable usage context?}  To address those
questions, we have the following design strategies.

{\bf S1. Pruning with Beam Search.} To deal with the scalability
issue of the exponential combination among all possible names of
variables, we use the Beam Search strategy: at a step during name
recovery, our algorithm keeps only the best $K$ sets of partially
recovered results according to the association scores.
%
%
%
This would help to reduce significantly the number of partially
recovered sets that need to be considered.

{\bf S2. Starting with Variable with Most Contextual Information.}
%
%
A naive answer is to use the appearance order~of the variables in the
code. However, the first variable in the code might not be the one
that we have sufficient information to recover its name. Thus, we
follow the idea of using~context to decide the initial variable
for name recovery. Our intuition~is that the more context information
a variable has, the more chance we have in correctly recovering its
name. {\tool} starts with the variable having
the most single-variable usage~information. That is, the variable has
the most relations with  method calls and
field accesses. The appearance order is used to break the tie if
multiple variables have the same number of relations in their
single-variable contexts.


{\bf S3. Selecting Next Variable with Greedy Strategy.} After one or
multiple variable names are recovered, we need to
determine which variable to recover next.
%
The appearing order in the code might not give us the optimal one. The
variable with the most single-variable context might not work either
since it might not go together well with others.
In {\tool}, we select the next variable $v$ with the list of
candidate names $vn$s that gives us the best partially recovered
result. That is, together with the selected names for the previously
recovered variables, the best possible choice for $v$ would give the
highest score with respect to the likelihood of the co-occurrences of
the recovered variables including $v$. This is a greedy strategy that
favors the variable and its candidate name that has most
co-appearances with the previously recovered variable names. It helps
avoid considering all possible candidate names for all the variables.
We use the phrase ``partially recovered result'' because only a subset
of all variables including $v$ is recovered for their names, while
other variables have not been processed.

%




\subsubsection{Detailed Algorithm}

\makeatletter
\def\BState{\State\hskip-\ALG@thistlm}
\makeatother



\begin{algorithm}[t]
	\caption{Multiple-Var Name Recovery Algorithm (\texttt{MVar})}
	\label{multiVarAlgorithm}
	\begin{algorithmic}[1]

		\Function{MVar}{$candidates[], N, Context$}
		\State $firstVar \gets \text{Pick the first var using } Context$
		\State $CList \gets candidates[firstVar]$
		\While{$nRecovered < N$}
			\State $nextVar \gets \text{Decide the next variable using S3}$
			\State $cl \gets candidates[nextVar]$
			\State $CList \gets BeamSearch(CList, cl)$
		\EndWhile
		\State $\textbf{Return } CList$
		\EndFunction
		
		\Function{BeamSearch}{$CList, cand$}
		\State $\textit{allPossiblePartialRecoveredSets} \gets \textit{CList} \otimes \textit{cand} $

		\For{\emph{partialRes} \textbf{in} \emph{allPossiblePartialRecoveredSets}}
		\State $MC$=$\textit{CalculateScore(partialRes)}$ via Formula~\ref{confidentScore}
		\EndFor
		
		\State $SortAll$ $(partialRes)$ \text{by} $MC$
		\State $TopRankedResult \gets TopKHighestScores$
		
		\State $\textbf{Return } TopRankedResult$
		\EndFunction
	\end{algorithmic}
\end{algorithm}

Algorithm~\ref{multiVarAlgorithm} shows the pseudo-code for our
algorithm, \texttt{MVar}, to derive names using the multiple-variable
contexts. Given a set of $N$ minified variables in which each variable
has a set of candidate names $candidates$ (provided by a $Context$,
\eg Single-variable or Task-specific contexts), \texttt{MVar}
determines the first variable to start. Using the strategy S2, it
chooses the first minified variable with the highest score according
to the context (line 2). For example, if the single-variable context
is chosen, Formula~\ref{eq:sgcs} is used.
%
All the candidates for the first variable are initially stored in the
current candidate list $CList$. Then, the algorithm iterates to recover the
variable names until all the variables are recovered (lines 4--7). At
a step of the iteration, assume that it has recovered $nRecovered$
variables. Using the strategy S3, the next variable $nextVar$ is
chosen such that together with the selected names for $nRecovered$
previously recovered variables, the possible names for $nextVar$ will
give the highest $MC$ score (Formula~\ref{confidentScore})
considering the co-occurrences of currently recovered variables
(lines 5--6).

After selecting the next variable, \texttt{MVar} performs beam search
by first generating all possible names for the ($nRecovered$ + 1)
variables with combining the name candidates for $nextVar$ and $CList$
(line 10). Each of those sets of names represents a partially
recovered result for those ($nRecovered$ + 1) variables. The scores of
all of those sets are computed (lines 11--12) using
Formula~\ref{confidentScore}. Then, we keep only the best $K$ sets of
results with highest scores and then store them in $CList$ (lines
13--14). \texttt{MVar} stops when all variables have been
recovered. $CList$ is returned as the list of best $K$ sets of
variable names for all the variables in the input (line 8).
%





In our example, using the \textit{single-variable context} and/or
the \textit{task-specific context}, we have the set of name candidates
for each variable, \eg
\texttt{r}:\texttt{(dataTransfer,dataContent,...)}, \texttt{f}:\texttt{(elementType,dataType,contentType,...)}, \texttt{p}:\texttt{(i, j,}...), \texttt{n}:\texttt{(data, cacheData, dataContent,...)}, etc. The
variable \texttt{r} is chosen first
since its single-variable context has most relations.
All the name candidates for \texttt{r} are stored in $CList$. After
that, the next variable is \texttt{n} because in all
the candidate names of the non-yet-recovered variables, the
name \texttt{data} for \texttt{n} when appearing with the current
candidate name of \texttt{r} gives the highest score. In this case,
the candidate name of \texttt{r} is \texttt{dataTransfer}. Then, all
the sets of variable names for \texttt{(r,n)}
include \texttt{(dataTransfer, dataContent), (dataTransfer, cacheData), (dataTransfer, data)}, etc.
Ranking these partially recovered results with
Formula~\ref{confidentScore} and keeping only the top $K$ ones,
$CList$ includes \texttt{(dataTransfer, data), (dataTransfer,
dataContent)},... The next variable for recovery
is \texttt{f}.
The process continues until all variables are recovered. $CList$
results are returned as the output.

\section{Task-specific context (TSC)}
\label{task_section}



In a program, a function has its functionality and is written to
realize a specific task. Each variable used in that function plays a
certain role toward that task.
%
Thus, the names of variables are relevant to the task
of the function and often consistent with one another.
The task of a function is typically described by a succinct name of
the function.
To derive a variable name using the task context, we use the
association relation to compute how likely a variable name appears
within a function with a particular name.
%
%
%
Given a variable name \textit{vn} and a function name \textit{fn},
{\bf Task Context score} (TC) represents the likelihood that the name
\textit{vn} appears within the body of a function named \textit{fn}.
We utilize fuzzy set theory~\cite{George:1995} as follows.
\begin{equation} \label{eq:fvas}
TC_{vn, fn}=\frac{N_{vn, fn}}{N_{vn} + N_{fn} - N_{vn, fn}}
\end{equation}
where $N_{vn, fn}$ is the number of functions in the corpus in which
\textit{vn} and \textit{fn} are observed together; $N_{vn}$ is the
number of functions in which \textit{vn} is used; and $N_{fn}$ is the
number of functions named \textit{fn}. As seen in
Formula~\ref{eq:fvas}, the value of TC is between [0,1]. The
higher the value $TC_{vn, fn}$, the higher the likelihood that the
variable name \textit{vn} appears in the function \textit{fn}.
In Figure~\ref{example_org}, using our experimental dataset, we can
compute $TC$ score between the function name
\texttt{getClipboardContent} and the variable name
\texttt{dataTransfer} as $TC$ = $\frac{3}{21 + 5 - 3} =
0.13$.

%

%
%
%

A function name might contain multiple tokens, \eg \texttt{get},
\texttt{Clipboard}, \texttt{Content} in
\texttt{getClipboardContent}. Each token contributes to an aspect to
emphasize the common task of the function. A variable might be
relevant to one specific aspect of the task. Thus, if we tokenize the
function names, we can account for those cases.
%
%
By tokenizing, a function name \textit{fn} could be represented by a
set of key tokens $S = \{t_1, t_2, t_3\dots\}$ (stopwords are removed)
and the $TC$ score between a variable name \textit{vn} and a function
name \textit{fn} is computed as follows:
\begin{equation} \label{eq:fvastoken}
TC_{vn, fn}=\max_{t \in S}{ \frac{N_{vn, t}}{N_{vn} + N_{t} - N_{vn, t}}}
\end{equation}
where $S$ is the set of key tokens of \textit{fn}; $t$ is a token in
$S$; $N_{vn}$ is the number of functions in which \textit{vn} is used;
$N_{t}$ is the number of functions containing token \textit{t}; and
$N_{vn, t}$ is the number of functions in which \textit{vn} and
\textit{t} occur together.



\section{Variable Name Recovery with Contexts}
\label{approach_section}




This section presents {\tool}, our approach to recover the variable
names in minified code using the combination of those above contexts.
%
Given a minified JS file, whose variables have been minified, {\tool}
produces a recovered JS file in which all variables are recovered with
meaning names.


Algorithm~\ref{JSNeatAlgorithm} shows the pseudo-code for {\tool}.
First, {\tool} builds the relation graph representing single-variable
context (SVC) for each variable in a function and derives the
candidate list. It then computes the task-specific context (TSC) for
each variable and derives the corresponding candidate list.
%
%
The two candidate lists produced by the contexts are combined into
a new list in which the likelihood that a candidate name
$vn$ is assigned to a variable $v$ is computed as~follows.
\begin{equation} 
\label{combinationSingle_Task}
ST_{v, vn} = \alpha \times  SC_{v, vn} + \beta \times TC_{vn, fn}
\end{equation}
where $fn$ is the function name; $TC_{vn, fn}$ is task context
score between $vn$ and $fn$; $SC_{v, vn}$ is the name
matching score between $vn$ and $v$, and $\alpha$ and $\beta$ are
weighting parameters, representing the importance of the contexts.




%

After this step, for each variable, we have a candidate list
($STList$) in which each candidate name has a score. {\tool} then uses
$STList$ as the input for \texttt{MVar} (Algorithm~\ref{multiVarAlgorithm})
to compute the top-ranked sets of names for all variables.





Note that the scoring function for a partially recovered result in
\texttt{MVar} algorithm (Formula~\ref{confidentScore}) needs to be
adjusted to account for the above combined score $ST$ as follows.
\begin{equation}
\label{confidentScoreV2}
  \begin{aligned}
MC_{vn_1, vn_2, ..., vn_n} = \gamma \times \frac{\sum_{j=1}^{n_{subJ}} assoc(S_j)}{n_{subJ}} \\
+ \theta \times \frac{ST_{v_1, vn_1} + ST_{v_2, vn_2} + ... + ST_{v_n, vn_n}}{n}
\end{aligned}
\end{equation}

where
$\gamma$ and $\theta$ are weighting parameters.

\begin{algorithm}[t]
	\caption{Context-based Name Recovery Algorithm}
	\label{JSNeatAlgorithm}
	\begin{algorithmic}[1]

		\Function{JSNeat}{\text{MinifiedFile} $f$}
		\State $SVCs \gets \text{Build Single-Var Contexts for variables in $f$}$
		\State $TSCs \gets \text{Compute TSC contexts for all variables in $f$}$
		\State $STList \gets \text{Combine cand lists from SVC and TSC}$
		\State $TopResults \gets$ \text{\texttt{MVar}} $(STList, N, SVCs)$
		\State $\textbf{Return } TopResults$
		\EndFunction

	\end{algorithmic}
\end{algorithm}



\section{Empirical Methodology}
To evaluate {\tool}, we answer the following questions:

\noindent \textbf{RQ1}:
\textit{Comparative Study}.
How accurate is {\tool} in name recovery for minified
JS code and how is it compared with the state-of-the-art
approaches, JSNice~\cite{JSNice2015} and
JSNaughty~\cite{JSNaughty2017}?

\noindent \textbf{RQ2}: \textit{Context Analysis}.
How do different combinations of contexts contribute to
{\tool}'s accuracy in different~settings?


\noindent \textbf{RQ3}: \textit{Sensitivity Analysis}.
How do various factors affect the accuracy, \eg
data's sizes, thresholds, parameters, etc.?

\noindent \textbf{RQ4}: \textit{Time Complexity}. What is {\tool}'s running time?

\vspace{-0.07in}
\subsection{Corpora}


We collected a corpus of 12,000 open-source JS projects from
GitHub with highest ratings.
For comparison, we followed the same procedure in previous
work~\cite{JSNice2015,JSNaughty2017} to collect
and clean up data.
We removed all duplicate files to avoid overlapping when testing. We
also removed the already-minified files because they will not help in
training.
Table~\ref{tab:corpusData} shows our dataset's statistics. As seen,
the number of unique variable names is much smaller than that of
variables. Thus, such repetition in names would help our data-driven
approach.

In our comparative study,
we used the same experimental setting as in JSNice~\cite{JSNice2015}
and JSNaughty~\cite{JSNaughty2017} by randomly splitting the dataset
into training and test corpora. In particular, to build {\bf Testing
  Corpus}, we randomly sampled 2K JS files in the dataset. The
remaining 320K files were used as {\bf Training Corpus}. The
level of sizes of testing and training data
(Table~\ref{tab:corpusData}) is comparable with that of the experimental
studies in existing tools~\cite{JSNice2015,JSNaughty2017}.
We minified the files using the minifying tool
UglifyJS~\cite{uglify}, and used the original files as oracle.

To build the relation graphs, we used Rhino to parse the JS files and
extract the context information. Table~\ref{tab:RGinfo} shows the statistics of our dataset $\mathcal{G}$ of relation graphs.


%

\begin{table}[t]
  \centering
  \caption{Data Collection}
    \begin{tabular}{c|c|c|c}
    \midrule
    \textbf{Category} & \textbf{Test Corpus} & {\bf Training Corpus} & {\bf Total} \\
    \midrule
    Files & 2K & 320K & 322K \\
    Functions & 6K & 961K & 967K \\
    Variables & 19K & 3481K & 3.5M \\
    Unique variable names & 5K & 171K & 176K \\
    Variables per file & 9.5 & 10.94 & 10.93  \\
    \bottomrule
    \end{tabular}%
  \label{tab:corpusData}%
\end{table}%

%
\begin{table}[t]
  \centering
  \caption{Database of Relation Graphs}
    \begin{tabular}{c|c}
    \midrule
    \textbf{Category} & \textbf{Quantity} \\
    \midrule
    Total number of graphs & 3.5M \\
    Mean number of graphs per file & 10.93 \\
    Mean number of graphs per function & 3.62 \\
    Min/Mean/Max number of edges per graph & 1, 2.2, 41 \\
    \bottomrule
    \end{tabular}%
  \label{tab:RGinfo}%
\end{table}%

\vspace{-0.06in}
\subsection{Evaluation Setup}


\subsubsection{Comparative Study}

For a tool under study, we trained~it with the Training Corpus and
tested it against the Testing Corpus. For JSNice, we used the
publicly~available tools on their website~\cite{jsnice-website} with
default parameters. For JSNaughty, we trained the translation and
language models, and Nice2Predict framework following the 
instructions. We did not run their tools in 10-fold cross validation
due to a long running time.






\subsubsection{Context Analysis}

We study the impact of different contexts, we created different
variants of {\tool} with various combinations of contexts and measured
their accuracies. We used the {\em 10-fold cross-validation setting}
on the entire corpus: 90\% of the files (9 folds) are used for
training and 10\% of the files (one fold) for testing, and we repeated
testing for each of the 10 folds and training with the remaining
folds. We also performed 10-fold cross validation on the project basis.



\subsubsection{Sensitivity Analysis}

To study different factors that have impact on {\tool}'s accuracy, in
our entire dataset, we randomly chose one fold for testing and the
remaining 9 folds for training. We studied the following factors:
relation graph size, type of relation edges, thresholds, beam sizes,
different degrees of associations, different weight parameters, and
data~size.

\subsection{Procedure and Metrics}

To measure the accuracy of a tool, we used
UglifyJS~\cite{uglify} to minify the given JS files,
and used the minified code as the input for the tool under study. We
then compared the resulting names from the tool against the original
names. Specifically, the tool is considered to correctly recover
the name of a variable $v$ if the recovered name $vn$ is matched
exactly with its original name. For $v$, if matching, we count it as
a hit, otherwise, it is a miss. Accuracy is measured by the ratio
between the total number of hits over the total number of~cases.





\section{Empirical Results}
\label{empirical_result_section}

\subsection{Accuracy Comparison (RQ1)}
\label{acc_comparison}

\begin{figure}[t]
	\begin{center}
	  \includegraphics[width=3.2in]{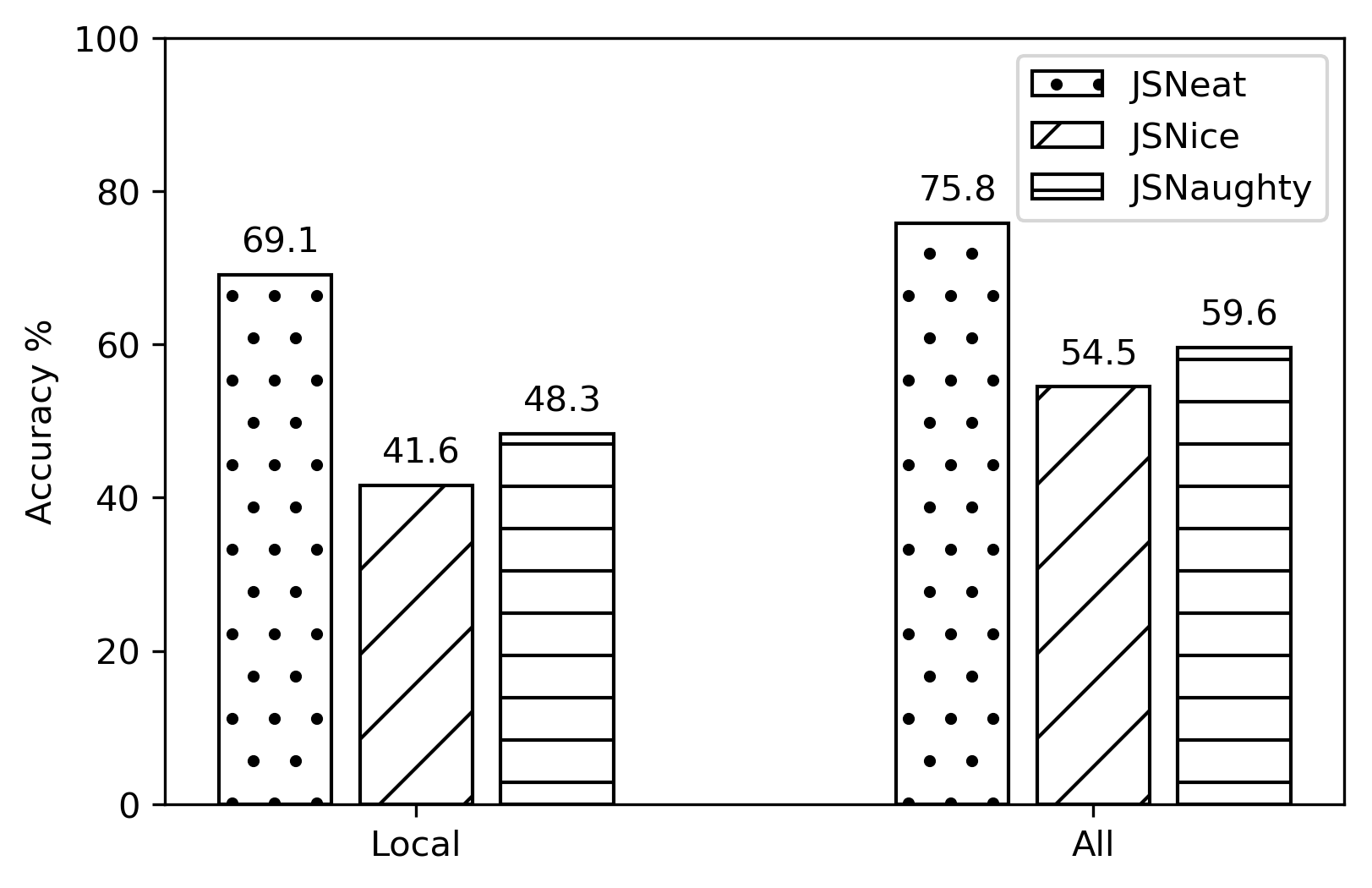}
          \vspace{-0.12in}
	\caption{Accuracy Comparison}
	\label{comparison_3tools}
	\end{center}
\end{figure}






In this study, we evaluate {\tool}'s accuracy and compare it with
JSNice~\cite{JSNice2015} and~JSNaughty~\cite{JSNaughty2017}.
As seen in Figure~\ref{comparison_3tools}, for~local variables,
{\tool} achieves high accuracy of {\bf 69.1\%}: relative improvements
of {\bf 66.1\%} and {\bf 43\%} over JSNice and JSNaughty,
respectively. The absolute improvements are 27.5\% and 20.8\%,
respectively.  For all variables (local and global ones), {\tool}
achieves even higher accuracy: {\em 75.8\%}, the relative improvements
of {\em 39\%} and {\em 27\%} over JSNice and JSNaughty,
respectively. The absolute improvements are 21.3\% and 16.2\%,
respectively.
Note that~global variables are not minified, we computed the accuracy
for~all variables for the completeness purpose. From now on, using~the
terms ``variables'', we refer to the recovery accuracy
for~local~variables.


\begin{figure}[t]
	\begin{center}
		\includegraphics[width=2.6in]{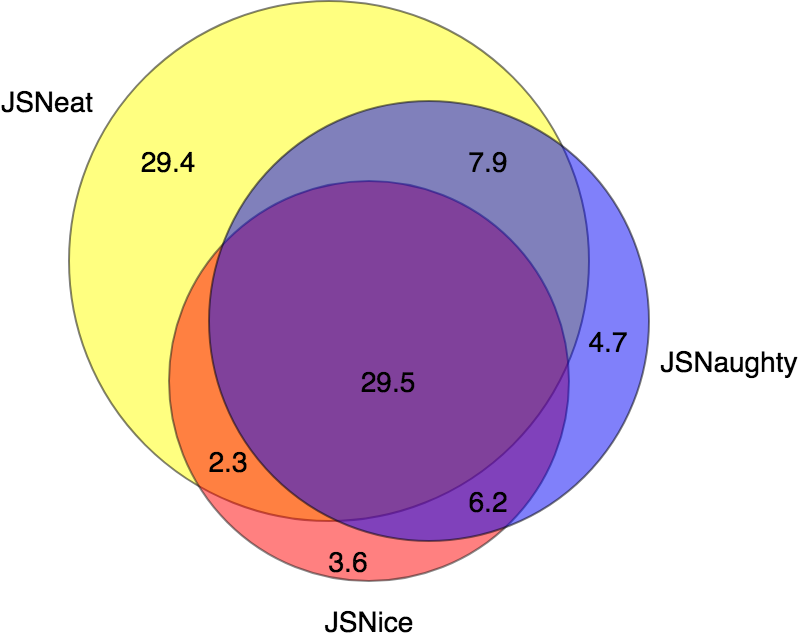}
		\caption{Overlapping among Results from the Tools}
		\label{overlapping}
	\end{center}
\end{figure}



%


We further analyze the overlapping between the results from three
tools.
The Vein diagram in Figure~\ref{overlapping} shows the percentages
of variable names that are correctly recovered.
As seen, a high percentage ({\bf 29.4\%}) of variables is
recovered only by {\tool}, while 4.7\% and 3.6\% of variables
are recovered only by JSNaughty and JSNice, respectively.
Meanwhile, there are 14.5\% of the variables that are correctly recovered
only by JSNice or JSNaughty, and not by {\tool}.



We also analyzed which contexts in {\tool} contribute to those 29.4\%
correctly recovered names. To do that, we {\em deactivated}
each of the three contexts.
%
%
%
{\em When MVC is disabled}, the percentage of variables that are 
correctly recovered only by {\tool} decreases by 7.4\% compared to the full
version.
%
%
However, {\em when TSC is disabled}, the accuracy drops dramatically from
29.4\% to 7.6\%.
%
Taking a deeper look in the cases of 29.4\%, we found out that 19\% of
variables that were recovered correctly when using all contexts become
incorrectly recovered ones when TSC is off.
83\% out of that 19\% come from the function with a single
variable. MVC certainly would not help in those cases because it needs
the contexts from other variables. In fact, in such functions, the
function names are quite relevant to the variable names. While SVC
does not have enough information to rank the correct name on the top,
TSC provides useful information to help in those cases.
For example, function \texttt{responseJson} in project
\texttt{mf-geoadmin3} uses a variable named \texttt{response}, that
was minified into \texttt{x}. Using only SVC and MVC, the correct name
\texttt{response} is ranked at position 5, but adding TSC,
\texttt{response} is ranked at the top. In the next experiment, we
{\em deactivated the single-variable context (SVC)}, and the percentage of
variables that are correctly recovered only by {\tool} decreases to
0.2\%. This means that SVC with property and role relations with the
pivots in the code contributes most to that 29.4\% of the
cases that were not recovered correctly by the other tools.

%


\subsection{Context Analysis Evaluation Results (RQ2)}
\label{intrinsic_section}

\begin{table}[t]
  \centering
  \caption{Impact of Contexts on Accuracy and Recovery Time}
  \label{combinations}
  \tabcolsep 4pt
	\begin{tabular}{||r | c || r | r||}
		\hline
		& {\bf Combination of Contexts}  & {\bf Acc (\%)} & {\bf Time (ms)} \\
		\hline
		1 & Task (TSC)  & 5.2 & 1.3 \\
		2 & SingleVar (SVC)   & 33.7 & 1.5 \\ 		
		3 & Task + SingleVar  & 47.3 & 2.1 \\
		4 & Task + MultiVar (MVC) & 9.3 & 2.3 \\
		5 & SingleVar + MultiVar & 37.8  & 2.6 \\
		\rowcolor{lightgray}
		6 & Task + SingleVar + MultiVar (= {\bf {\tool}}) & 63.1 & 3.2 \\
		\hline
	\end{tabular}
	
\end{table}


As seen in Table~\ref{combinations}, using only task-specific context~(TSC) (line 1), 
accuracy is low because all the variables in~the~same
function have the same chance to be recovered with a certain name. In
contrast, the single-variable context (SVC)~achieves much higher
accuracy (33.7\%). This is reasonable since SVC provides more detailed
context for individual~variables such as the relations to surrounding
method calls and field~accesses.

Combining TSC and SVC provides an additional improvement of 13.4\%
over the tool with only SVC (lines 2 and~3). We found that several
correct candidate names that were ranked in the 2nd-4th positions
become the top candidates~with the addition of TSC. In contrast, the
combination of TSC and MVC yields only slight improvement over TSC (5.2~to~9.3). 
The reason is that MVC takes the lists of candidate names as
its input and such lists were not initially of high quality (only
5.2\% accuracy), leading to low accuracy. For the combination of SVC
and MVC, the improvement is 4.1\% over SVC (lines 2 and 5). We found
that those 4.1\% of cases, the co-occurrences of variable names help
rank them in the top~positions. Comparing lines 3 and 5, adding TSC to
SVC improves almost 10\% more than adding MVC to SVC. Further
analyzing, we found that such improvement from TSC is for the cases in
which 1) the given JS function has only one variable (MVC cannot help)
and/or 2) the SVC has only one relation (SVC did not perform well with
little surrounding context).

Finally, combining three contexts, {\tool} achieves the highest
accuracy. Compared to TSC+SVC (lines 3 and 6), {\tool}
relatively improves {\bf 33.4\%} (15.8\% absolute improvement). This
is reasonable since the two contexts TSC and SVC alone achieve the highest
accuracy among all the combinations of two contexts. Thus, they give
MVC algorithm the initial candidate lists for variables with higher
quality.~Then, MVC with the co-occurrence information among variables
helps an additional improvement of 15.8\%. Moreover, comparing lines 5 and
6, TSC helps improve 25.3\% since TSC helps in the cases of
single-variable functions or single-edge~RGs.

To evaluate {\tool}'s consistency in achieving high accuracy, we
performed 10-fold cross validation.  As seen in~Table
\ref{run10folds1}, the accuracies for all the folds are stable
(61.9--63.8\%), with the recovery time of 2.92~ms for a file.  The
results~for 10-fold cross validation on project basis are similar
(not~shown).



\begin{table}[t]
  \centering
  \caption{10-fold Cross-Validation Evaluation on {\tool}}
	\label{run10folds1}
  \scriptsize
  \tabcolsep 3.2pt
	\begin{tabular}{||c || c | c | c | c | c | c | c | c | c | c || c||}
		\hline
		Test Fold & 0 & 1 & 2 & 3 & 4 & 5 & 6 & 7 & 8 & 9 & Mean \\
		\hline
		Acc (\%) & 63.1 & 63.0 & 63.1 & 61.9 & 63.3 & 62.7 & 63.8 & 63.4 & 62.9 & 62.5 & 62.9 \\
		Time & 2.9 & 2.92 & 2.86 & 2.83 & 2.95 & 2.99 & 2.92 & 2.95 & 3.01 & 2.85 & 2.92 \\
		\hline
	\end{tabular}
\end{table}

\subsection{Sensitivity Analysis Evaluation Results (RQ3)}
\label{sensitivity_section}

\vspace{-0.018in}
\subsubsection{Impact of Relation Graphs' Sizes}


\begin{table}[t]
	\centering
        \caption{Impact of Relation Graphs' Sizes on Accuracy}
	\label{analysisByEdge}
        \tabcolsep 5.5pt
	\begin{tabular}{||c || c | c| c| c| c| c| c||} 
		\hline
		Number of edges & 1 & 2 & 3 & 4 & 5 & >5 & All\\
		\hline
		\% of graphs  & 46.4 & 24.8 & 12.8 & 6.6 & 3.5 & 5.7 & 100 \\
		\hline
		Accuracy (\%) & 58.5 & 65.7 & 67.4 & 68.4 & 68.7 & 69.8 & 63.1 \\
		\hline
	\end{tabular}
\end{table}


To measure the impact of the sizes of SVC, we selected in the corpus
only certain sizes of RGs (measured by the number of edges).
As seen in Table~\ref{analysisByEdge}, the more relations in SVC
to be considered (more edges a relation graph has), the more
accurate a variable name can be recovered. When having more than 4
edges, accuracy becomes stable at a high level and gradually
increases.  This also reaffirms our strategy S1 in selecting the first
variable with most connecting edges in RGs.




\subsubsection{Impact of Type of RG Edges}

By omitting only certain type of edges in RGs for SVC,
we measured the impact of each type of
relations on accuracy.
In Table~\ref{analysiByRelation}, the lower the accuracy, the higher
the impact the corresponding relation has.
As seen, all the relation types contribute nearly equally to
{\tool}. If one type of relation is not considered, accuracy drops
from 63.1\% to around 45\%. The \texttt{argument} relation has a
slightly higher contribution than the others.


\begin{table}[t]
	\centering
        \caption{Impact of Relation Types in RGs on Accuracy}
	\label{analysiByRelation}
        \tabcolsep 5pt
	\begin{tabular}{||c || c | c | c | c||} 
		\hline
		& argument & assignment & fieldAccess & methodCall \\
		\hline
		Accuracy (\%)  & 42.9\% & 44.5\% & 48.5\% & 45.7\% \\
		\hline
	\end{tabular}
\end{table}

\subsubsection{Impact of Threshold for Graph Matching}

\begin{table}[t]
	\centering
        \caption{Impact of Threshold $\varphi$ on Accuracy}
	\label{thresholdSensitivity}
	\begin{tabular}{||c || c | c | c | c | c | c||} 
		\hline
		Threshold $\varphi$ & 0.5 & 0.6 & 0.7 & 0.8 & 0.9 & 1.0 \\
		\hline
		Accuracy (\%) & 27.1 & 27.4 & 30.6 & \textbf{33.7} & 31.2 & 30.8 \\
		\hline
		Time &  1.8 & 1.7 & 1.5 & 1.5 & 1.4 & 1.4 \\
		\hline
	\end{tabular}
\end{table}


We evaluated the impact of the threshold $\varphi$ used to measure
the similarity between two RGs (\ie two SVCs).
%
To do that, we used only SVC
to recover variable names with varied $\varphi$  [$0.5$--$1.0$]. As
seen in Table~\ref{thresholdSensitivity}, when $\varphi$=$0.8$, the
accuracy is at the highest.
With $\varphi <$ 0.8, the number of variables whose contexts are
matched with the minified variable is large, and the correct name was
not ranked at the top. When $\varphi >$ 0.8, the condition is too
strict and the correct names were dropped because it is not easy to
find a completely matched context.

\subsubsection{Impact of Beam Size}
\label{beam_size_section}

\begin{figure}[t]
	\begin{center}
		\includegraphics[width=3.0in]{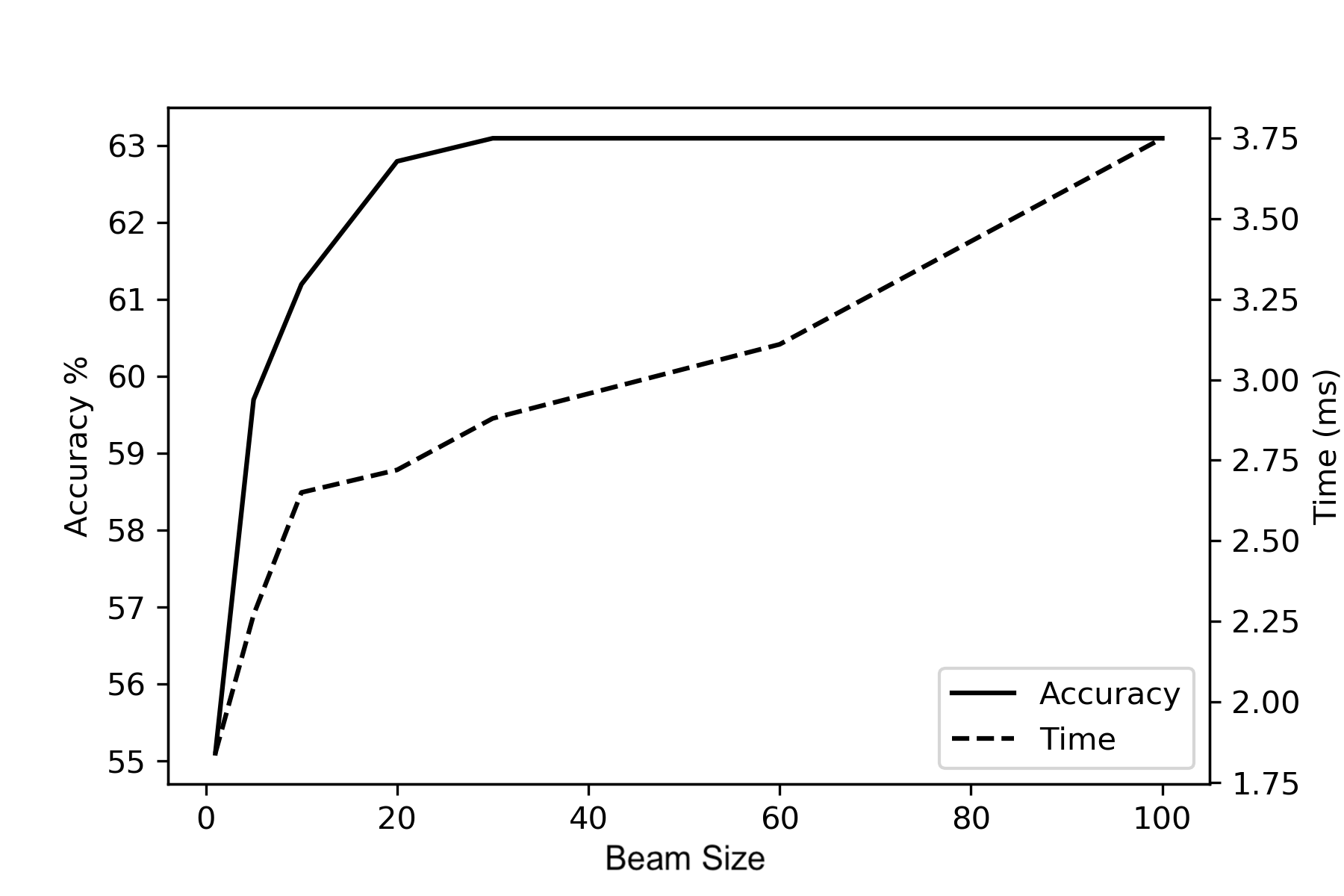}
		\caption{Impact of Beam Size on Accuracy and Running Time}
		\label{beamwithvsAccuray}
	\end{center}
\end{figure}


In {\tool}, beam size is used to deal with the large combinations of
possible names.
Figure~\ref{beamwithvsAccuray} shows the accuracy and running time per
variable when we varied the beam sizes.  As seen, when the beam size
is small, the accuracy is very low. It is expected because pruning
occurs frequently, the number of
results that were kept is smaller, and the best candidates might be
dropped out of the beam stack. As the beam size is increased, accuracy
increases and reaches the highest point (around 63\%) with the beam
size of 30. Accuracy becomes stable when the beam size is greater than
30. The reason is that almost all the correct names are observed in
the top 30 results. Therefore, when we increase beam size over 30,
accuracy is not affected anymore. Regarding the running time, the
higher the beam size, the larger the number of candidate results, and
the higher the running time. Thus, we used 30 for the beam size in
other experiments.




\subsubsection{Impact of Association Scores}
\label{assoc_section}


\begin{table}[t]
        \centering
        \caption{Impact of Assoc Score $J$ on Accuracy and Time}
        \label{associationSensitivity}
        \begin{tabular}{||c || c |c |c ||}
                \hline
                 $J$ & Accuracy (\%) & Time (ms)& Perc. Found (\%)\\
                \hline
                Pairwise ($J=2$) & 63.1 & 2.88 & 68.5\\
                \hline
                Triple ($J=3$) &  61.8 & 3.11 & 55.2\\
                \hline
                $J=4$ &  60.2 & 3.26 & 25.7\\
                \hline
                $J=6$ &  58.6 & 3.41 & 14.9\\
                \hline
                $J=8$ &  57.5 & 3.45 & 5.8\\
                \hline
                $J=10$ &  57.4 & 3.47 & 2.7\\
                \hline
        \end{tabular}
   \end{table}


In MVC, {\tool} considers the co-occurrences of $J$ = 2, 3, etc or
$n$ variable names.
%
The value of $J$ shows how many associations of variable names
that we need to have high accuracy.
In~Table~\ref{associationSensitivity}, using high-degree
association, accuracy decreases gradually since finding
co-occurrences of $n$ variable names has a lower probability than
finding the co-occurrences of $n$-1 variable names, \ie the
co-occurrence condition is too strict. The decrease in accuracy
is not much since the high-degree association affects only a smaller
set of cases with higher numbers of variables in a function. As
expected, running time~increases.




\subsubsection{Impact of Parameters in Context Combinations}
\label{combine_sensitivity}

To combine multiple contexts, we use parameters to put weights on
each of them, \eg $\alpha$ for SVC, $\beta$ for TSC, and
$\gamma$ for MVC. We varied their values
to observe the impacts of the contexts.
In Table~\ref{params}, when combining SVC and TSC, if the
weight of SVC is higher, accuracy is higher.
In a function, task context plays an equal role to all variables in
that function, while SVC provides directly related information to the
variable.
%
%
Combining SVC, TSC, and MVC, the higher $\gamma$, the
higher the accuracy. This means that MVC contributes more important
information than SVC and TSC in the Formula~\ref{confidentScoreV2}.

\begin{table}[t]
	\centering
        \caption{Sensitivity Analysis on Combination Parameters}
	\label{params}
		\begin{tabular}{||c | c | c || c | c | c ||}
			\hline
			$\alpha$ &	$\beta$	& Accuracy (\%)	& $\theta$ &	$\gamma$ &	Accuracy (\%)\\
			\midrule
			1 & 0 & 45.4 & 1 & 0 & 57.9 \\
            0.25 & 0.75 & 45.9 & 0.75 & 0.25 & 58.5 \\
			0.5	& 0.5	& 46.5	& 0.5	& 0.5	& 60.2 \\
            0.75  & 0.25 & 47.3 & 0.25 & 0.75 & 61.7 \\
			0 & 1 & 46.7 & 0 & 1 & 63.1 \\
			\hline
		\end{tabular}
\end{table}


\subsubsection{Impact of Training Data Size}
\label{data_size}

\begin{table}[t]
	\centering
        \caption{Impact of Training Data’s Size on Accuracy}
	\label{run10folds}
\tabcolsep 4.2pt
		\begin{tabular}{||c || c | c | c | c | c | c | c | c | c ||}
			\hline
			\#Folds & 1 & 2 & 3 & 4 & 5 & 6 & 7 & 8 & 9\\
			\hline
			Acc.Full & 40.1 & 47.3 & 52.5 & 54.9 & 56.4 & 58.2 & 60.0 & 62.2 & 63.1  \\
			\hline
			Acc.Token & 42.3 & 48.5 & 53.6 & 55.1 & 56.5 & 57.3 & 57.9 & 58.5 & 59.1 \\
			\hline
		\end{tabular}
	
\end{table}



To measure impact of data size, we used one fold for testing and
increased the sizes of the Training dataset by adding one fold at a
time until~9 remaining folds are added for training. We ran {\tool} on
each training dataset with the best settings for SVC and MVC, and two
settings of TSC: tokenizing or using full~function~names.~In
Table~\ref{run10folds}, the accuracies in both settings increase
linearly and consistently with the training size.
%
With a small size, tokenizing function names gives better accuracy
than using full names. However, when the number of training
folds is more than 5 folds, using full function name is better. The
reason is that when the data's size is large enough, the
probability of co-appearances between a function name and a variable
name is higher, the candidate names for a variable can be found
better. When data's size is small, {\tool} might not see a
variable name and a function name appearing together, then tokenizing
function names would give more useful context.

\vspace{-0.02in}
\subsection{Time Complexity (RQ4)}
\label{time_section}

\begin{table}[h!]
	\centering
        \caption{Running Time Comparison}
	\label{timeCompare}
	\begin{tabular}{||c || c | c | c||}
		\hline
		Metric  & \tool & JSNice & JSNaughty \\
		\hline
		Training & 2h05m & 8h35m & 12h25m \\
		Per-file Recovery  & 32ms & 72ms & 129ms \\ 		
		Per-variable Recovery & 2.9ms & 6.6ms & 11.8ms \\
		\hline
	\end{tabular}
\end{table}

All experiments were run on a Linux server with 20 Intel Xeon 2.2GHz
processors, 256GB RAM. In Table~\ref{timeCompare}, the time to recover
for a file or for a variable with {\tool} is {\em twice as fast as
  with JSNice and 4x as fast as with JNaughty}. More importantly, {\em
  {\tool}'s training time is 4x faster than JSNice and 6x faster than
  JSNaughty}.  This can be achieved due to the nature of information
retrieval in {\tool}, in comparison to machine learning in JSNice and
JSNaughty.



\subsection{Limitations and Threats to Validity}

\subsubsection{Limitations} First, as a data-driven approach, unseen
data affects our accuracy. For example, with 1-fold training data,
33.4\% of minified names have not been observed. Second, for the task
context, {\tool} did not work well for functions with general names,
\eg \texttt{next}, \texttt{find}, etc. More sophisticated solution
could involve topic modeling~\cite{LDA} on the function body. Third,
for MVC, greedy strategy might not achieve the optimal
result. Finally, if two variables in the same function are assigned
with the same name (\eg same SVC, MVC, and TSC), we randomly pick
different names. Program analysis could be applied here to improve
accuracy.

\subsubsection{Threats to Validity}

Our corpus of JS code might not be representative, however, we chose a
large corpus with the size comparable with those in previous
studies. We used only the tool Uglify to minify the code, which was also used
in JSNice and JSNaughty. We do not study the usefulness involving
human subjects. However, for comparison, we used the same experimental
settings as in JSNice~\cite{JSNice2015} and
JSNaughty~\cite{JSNaughty2017}.

\section{Related Work}
\label{related_section}

{\tool} is closely related JSNice~\cite{JSNice2015} and
JSNaughty~\cite{JSNaughty2017}.
JSNice~\cite{JSNice2015}
uses the graph representation of variables and
surrounding program entities via program dependencies. It infers
the variable names as a problem of structured prediction with
conditional random fields (CRFs)~\cite{JSNice2015}. In comparison,
first, while JSNice uses ML,
{\tool} is IR-based in which
it searches for a list candidate names in a large code corpus. Second,
{\tool} considers not only the impacts of surrounding program
entities in SVC, but also task and multiple-variable contexts.
Third, with CRF, JSNice is effective when variables have more
dependencies, and less effective~with the functions having one
variable. Finally, {\tool} is much faster and
the results are more accurate as shown in
Section~\ref{empirical_result_section}.

JSNaughty~\cite{JSNaughty2017} formulates name recovery as a
statistical machine translation from the minified code to the
recovered code. First, due to the nature of ML, it faces the
scalability issue in much higher time complexity. Second, JSNaughty
uses a phrase-based translation model, which enforces a strict order
between the recovered variable names in a function. This is too
strict since a name of a variable might not~need~to occur before
another name of another variable.
Third, JSNaughty does not consider the task context of
the variables.
Finally, our training/testing time is much faster.
In contrast, other~deobfuscation methods use static/dynamic
analyses~\cite{Christodorescu:2003:SAE:1251353.1251365,Moser:2007:EME:1263552.1264210,Udupa05deobfuscation:reverse}.

Statistical NLP approaches have been used in SE.
%
Naturalize~\cite{barr-codeconvention-fse14}  enforces a
consistent naming style.
%
%
Other applications of statistical NLP include code
suggestion~\cite{hindle-icse12,tbcnn14}, code
convention~\cite{barr-codeconvention-fse14}, method name
suggestion~\cite{sutton-fse15}, API suggestions~\cite{raychev-pldi14},
code mining~\cite{sutton-msr13}, type resolution~\cite{icse18}, pattern mining~\cite{sutton-16}.
Statistical NLP was used to generate code from text, \eg
SWIM~\cite{Raghothaman-ICSE16}, DeepAPI~\cite{gu-fse16},
Anycode~\cite{anycode-oopsla15}, etc.



\section{Conclusion}

This paper presents {\tool}, an IR-based approach to recover the
variable names in minified JS code. We follow a data-driven
approach
by searching for names in a large corpus of open-source JS code. We use
three types of contexts to match a variable in given minified code
against the corpus.
Our IR approach enables us to achieve
high accuracy with less time complexity than the state-of-the-art
approaches.
{\tool} achieves a high accuracy of 69.1\%: the improvement of 66.1\%
and 43\% over JSNice and JSNaughty, respectively.
The time to recover
for a file or for a variable with {\tool} is twice as fast as with
JSNice and 4x as fast as with JNaughty, respectively.


\section*{Acknowledgment}
This work was supported in part by the US National Science
Foundation (NSF) grants CCF-1723215, CCF-1723432, TWC-1723198,
CCF-1518897, and CNS-1513263.



%




\newpage

\balance

\bibliographystyle{abbrv}
\bibliography{icse19}




\end{document}